\documentclass[11pt]{article} 
\usepackage{epsfig}
\usepackage{color,cite,jcappub}
\newcommand{\Fig}[1]{Fig.\ \ref{#1}}
\def\Eqn#1{Eq.\ (\ref{#1})}

\def\cn{\mathop{\rm cn}}
\author[a]{Arindam Mazumdar,}
\affiliation[a]{Theory Division, Saha Institute of Nuclear Physics, Kolkata-64, India}
\emailAdd{arindam.mazumdar@saha.ac.in}
\author[b]{Kamakshya Prasad Modak}
\affiliation[b]{Astroparticle Physics and Cosmology Division, Saha Institute of Nuclear Physics, Kolkata-64, India}
\emailAdd{kamakshya.modak@saha.ac.in}
\date{}
\abstract{
We present a framework for calculating super-horizon curvature perturbation from the dynamics of
preheating, which gives a reasonable match to the lattice results. Hubble patches with
different initial background field values evolve differently. From the bifurcation of their
evolution trajectories we find curvature perturbation using Lyapunov theorem and $\delta N$
formulation. In this way we have established a connection between the finer dynamics of preheating and
the curvature perturbation produced in this era.
From the calculated analytical form of the curvature perturbation we have derived 
the effective super-horizon curvature perturbation smoothed out on large scales of
CMB. The order of the amount of local form non-gaussianity generated in this process has been calculated and 
problems regarding the precise determination of it have been pointed out.}

\begin{document}

\title{\bf Deriving super-horizon curvature perturbations from the dynamics of preheating }
\maketitle
\section{Introduction}
Preheating is expected to be a mechanism for producing
non-gaussian curvature perturbation in super-horizon
scales~\cite{Enqvist:2005qu,Enqvist:2004ey,Bartolo:2003gh,Kohri:2009ac,
Chambers:2008gu, Chambers:2007se,Jokinen:2005by,Byrnes:2008zz,Liddle:1999hq}. 
It is agreed in consensus that during preheating highly nonlinear 
and non-gaussian curvature perturbations generate in small 
scale~\cite{Bond:2009xx,Frolov:2010sz,Ackerman:2004kw}. But if it is asked  
that how much effect on the large scales of Cosmic Microwave Background(CMB)
can be generated by preheating, the answers vary in a wide range. 
To predict if the curvature perturbation produced during preheating era can contribute substantially 
one needs to know the functional form of curvature perturbation in small
scales~\cite{Suyama:2013dqa}. Here we present a formulation to analytically
calculate curvature perturbation from preheating which gives a good match with
lattice results\cite{Bond:2009xx}. This allows us to establish a connection between the different types of
processes during preheating and curvature perturbation produced due to them. Finally we 
derive the analytical form of effective curvature perturbations on large scales 
and estimate the amount of non-gaussianity arising from it.

The topic of super-horizon curvature perturbation and local form non-gaussianity 
produced during preheating has been addressed in many different ways in the literature.
Evolution of first order curvature perturbation was observed in \cite{Hamazaki:1996ir,
Bassett:1999cg, Finelli:2000ya,Zibin:2000uw}
for the matter fields having solutions of the form of reheating or preheating era. 
Later the authors of Ref.~\cite{Enqvist:2004ey,Enqvist:2005qu} solved
Einstein's equation with second order curvature perturbation for the
period of preheating, and took the limit $k\rightarrow 0$ of the solution. A high value
of non-linearity parameter $f_{\rm NL}$ was reported to be coming from preheating in different 
models~\cite{Enqvist:2005qu, Jokinen:2005by}. But there are some restrictions of this 
formulation, since it is neither a fully nonlinear approach to calculate preheating 
contribution, nor the gauge invariant combination of perturbations are well defined  
in second order~\cite{Bruni:1996im,Christopherson:2014bea, Dias:2014msa}. To avoid this problem
very recently covariant formulation has been used in deriving amplification of entropic 
modes during preheating period~\cite{Moghaddam:2014ksa}. There is another approach which 
uses ``separate universe approximation'' or $\delta N$ formulation~\cite{Bartolo:2003gh,
Kohri:2009ac,Chambers:2008gu, Chambers:2007se}. 
In this formalism first curvature perturbation in small scales was calculated using 
lattice simulation or other methods. After that it was smoothed out with a suitable
Gaussian window function and then its effect on CMB was predicted. We follow this second 
approach in our work. 

Separate universe approximation or the $\delta N$ formulation was first used in the
context of preheating in Ref~\cite{Tanaka:2003cka,Suyama:2006rk}. This formulation tells that the
curvature perturbation on super horizon scales can be written in terms of the difference 
in the number of $e$-foldings($N$) of different causally disconnected patches of the universe
on constant density  gauge\cite{Sugiyama:2012tj,Lyth:2004gb}. And the 
dependence of $\delta N$ with fields' perturbation on the super-horizon scales can be 
written as\cite{Lyth:2005fi},
\begin{eqnarray}\label{zetaexpandx}
 \zeta(x)=\delta N(x)={\partial N\over\partial \phi_I}\delta\phi_I(x) +{\partial^2 N\over\partial \phi_I\partial\phi_J}\delta\phi_I(x)\delta\phi_J(x). 
\end{eqnarray}
Therefore, if one can calculate the dependence of the $N$ with respect to the background field values
$\phi_I$s, curvature perturbation can be calculated in terms of field perturbations. Keeping 
this prospect in mind first lattice simulation to calculate $\delta N$ for different background 
values of the secondary field(say $\chi$) during preheating was attempted by the authors of
Ref.\cite{Chambers:2008gu,Chambers:2007se}. But the correct dependence was shown in Ref.\cite{Bond:2009xx}, 
where the authors found a ``periodic'' and ``spiky'' behavior in $\delta N(\chi)$. 
Even for some values of $\chi$ the height of those spikes are so large that the authors 
moved on to  postulate that ``the cold spot'' of CMB might have come from the dynamics of preheating. 

But even after all these efforts in this direction one thing that could not be concluded
properly is the amount of non-gaussianity preheating can produce in CMB. So, when
Planck released its data~\cite{Ade:2013ydc}, all other mechanisms of producing primordial
non-gaussianity got constrained, but nothing could be said about preheating. The
seed of the uncertainty in this question lies in the functional form of the
curvature perturbation. Lattice result gives spiky pattern, but that pattern can
be fitted with a large variety of functional forms and someone cannot logically
pick a particular one. Different assumption of functional form can give very different result of $f_{\rm
NL}$\cite{Suyama:2013dqa}. Therefore in this paper we attempt to give a functional form of curvature
perturbation from the basic dynamic of preheating, so that this uncertainty can
be resolved.         

The model of our consideration is
\begin{eqnarray}\label{potential}
V={\lambda\over 4}\phi^4+{g^2\over 2}\phi^2\chi^2 \, ,
\end{eqnarray}
since its lattice simulation is already available in literature and for small
value of $g^2\over \lambda$ it has been predicted to provide detectable
non-gaussian signature in CMB\cite{Bond:2009xx}. 
Although $\lambda\phi^4$ potential has been observationally ruled out to be a candidate 
of inflation potential\cite{Planck:2013jfk}, studying this model for preheating era is still important. It is 
because some possible potential form for small field inflation can also be approximated as $\lambda\phi^4$ during
the oscillation of $\phi$ around minima\cite{Felder:2001kt}. That means the Klein-Gordon equation for $\chi$ 
would have same mathematical form as in the $\lambda\phi^4$ potential.

Difference in the number of $e$-foldings($\delta
N$) in different causally disconnected patch of the universe with different
initial background values of $\chi$ (say $\chi_i$) gives curvature
perturbation.    
$\delta N(\chi_i)$ is observed at a fixed time sufficiently after the end
of preheating. Different values of $\chi_i$ drive the Hubble parameter $H(t)$ differently, and
difference in $\int H(t) dt$ gives the $\delta N(\chi_i)$. Since $\delta N$ formulation
is defined in constant density gauge we have to evaluate $\int H(t) dt$ up to the 
time when that particular patch reaches a fixed value of $H$. If $t_e^0$ and $t_e$ are
the required time for the disconnected patches with $\chi_i=0$ and and $\chi_i$ nonzero, to reach $H$ 
we have
\begin{eqnarray}\label{deltaN}
 \delta N(\chi_i)= \int_0^{t_e}H(\chi_i,t) dt -\int_0^{t_e^0}H(0,t) dt \, \, .
\end{eqnarray}
 So, $\delta N(\chi_i)$ can be evaluated if we can calculate how does different
$H(t)$ trajectories with different initial conditions move away or towards each other.
 
 We divide this paper in the following sections. In section \ref{rev-preheating}
and \ref{rev-preheating} we briefly describe the necessary concepts required
for the calculations done in the next sections. In section \ref{genform} we
will describe how can we calculate $\delta N$ from the bifurcation of the Hubble
trajectories. Then in the section \ref{dN-vs-gl} we will show that we can assume
some particular form of potential and kinetic energy terms and using those we can
calculate $\delta N$ for certain choices of $g^2$ and
$\lambda$. Comparison of lattice result and our expression will be shown in that section.
In section \ref{smoothing} we smooth out the $\delta N$ on large
scale. This requires a particular value of spatial variance of $\chi$ field
which corresponds to a certain scale of interest. The smoothed out function
gives us the necessary functional form of $\delta N$ required to calculate
$f_{\rm NL}$. After calculating $f_{\rm NL}$ we move on to the following section
\ref{conclusion} where we discuss the findings and the possible outcome of our
research. Necessary calculations and used techniques are described in three
appendices\ref{nodeshiftcal}, \ref{deltaNcalculation} and \ref{Edgeworthexpansion}.    
 
\section{Process of preheating}\label{rev-preheating} 
Before we begin describing the method for calculating $\delta N$, we will very briefly 
review process of preheating in this section which will be required to develop the
discussions in next sections.
\subsection{Parametric resonance} 
After the end of inflation inflaton $\phi$ 
oscillates around the minima of the potential and this oscillation pumps up the
creation of $\chi$ particle. For the potential of \Eqn{potential} equation of
motion for the classical background $\phi(t)$ becomes
\begin{eqnarray}\label{eom1}
\ddot\phi+3 H \dot\phi+\lambda\phi^3=0 \ .
\end{eqnarray}
Here dot means derivative w.r.t time($t$). This equation gives a solution of
$\phi(t)$ which oscillates but its amplitude decays with time. So one moves to
use
conformal
variables $d\eta= {dt\over a}, \Phi(\eta)=a\phi(\eta)$ and
$X(\eta)=a\chi(\eta)$. This
allows one to write \Eqn{eom1}
\begin{eqnarray}
{\partial^2\Phi\over\partial\eta^2}+\lambda\Phi^3 =0 \ .
\end{eqnarray}
Unlike the $m^2\phi^2$ potential in
this case the $\Phi$ does not oscillates sinusoidally, but the solution of this
equation can be written as $\Phi(\eta)=\tilde\Phi f(x)$, where
$x=\sqrt{\lambda}\tilde\Phi\eta$,
$\tilde\Phi$ is the amplitude of oscillation and $f(x)$ is given by Jacobi elliptic
function as 
\begin{eqnarray}
 f(x)= \cn\left(x-x_0,{1\over\sqrt{2}}\right)\ .
\end{eqnarray}
Elliptic function $f(x)$ can be expressed as a series of cosine functions. The
leading order term  is $c_0\cos(0.8472 x)$ where $c_0\approx 1$. Amplitudes of
the higher order terms fall off quite fast. The oscillation of classical field $\phi$ 
drives the parametric resonance of the quantum field $X(\eta)$. 
The mode equation of the $k$-th mode of $X(x)$ takes the following form 
\begin{eqnarray}\label{LameeqnX}
 X_k''+\left(\kappa^2+{g^2\over\lambda}{\cn}^2\left(x,{
1\over\sqrt{2}}\right)\right)X_k=0 \ .
\end{eqnarray}
Here derivative has been taken w.r.t. $x$ and $\kappa^2 ={k^2\over\lambda\tilde\Phi^2}$. 
This equation is a type of Lame equation. The solution of this equation grows as 
$X_k\propto e^{\mu_k x}$, where $\mu_k$ is the characteristic exponent of the corresponding 
$k$ mode. Value of $\mu_k$ depends on the value of $g^2\over\lambda$ and $k$. 
Maximum value it can attain is 0.2377. From the stability-instability 
chart of \cite{Greene:1997fu, Frolov:2010sz} one can figure out that for
${g^2\over \lambda}=2$, $\mu_k$ reaches maximum value for $k=0$
mode. 

Mode equation for $\Phi$ becomes 
\begin{eqnarray}\label{Lameeqnphi}
 \Phi_k''+\left(\kappa^2+ 3 {\cn}^2\left(x,{
1\over\sqrt{2}}\right)\right)\Phi_k=0 \ .
\end{eqnarray}
This equation is nothing but a particular case of \Eqn{LameeqnX}. Maximum growth of 
the inhomogeneous modes of $\Phi_k$
occurs for $\kappa=1.6$.

We will use these solutions as inputs in our calculation of $\delta N$ in section 
\ref{dN-vs-gl}.

\subsection{End of preheating}\label{endofpreheating}
Second stage of preheating when the amplification of $\chi$ and $\phi$ field affect
the parametric resonance, is known as back-reaction. Thus the frequency of the $\Phi$
oscillation increases significantly and the amplitude of the $\Phi$
decreases. In the first phase of preheating the the number density of the $\phi$ particle, 
say $n^{\phi}_k$ increases and the drains away the energy from the $\Phi$ field. Increase 
of $\langle \delta\phi^2 \rangle$ can happen in two ways. First one is 
parametric resonance described above and second one is rescattering.

Rescattering is the process of production of $\phi$ particles from the $g^2\phi^2\chi^2$ 
interaction term. Since the number density of $\chi$ particle increases exponentially this
process becomes more effective at the last stage of preheating. In fact the parametric amplification
of $\langle \delta\phi^2 \rangle$ is much less efficient than the rescattering effect. 

Amplitude of $\langle \delta\phi^2 \rangle $ is proportional to the squared of the number 
density of $\chi$ particle (say $n^{\chi}_k$), where as $n^{\chi}_k$ is proportional to 
the $\langle \chi^2 \rangle $. This means $\langle \delta\phi^2 \rangle $ $\propto\exp(4\mu x)$.  
This amplification of $\langle \delta\phi^2 \rangle $ can affect the parametric resonance in 
two ways. If the leading mode in $\langle \delta\phi^2 \rangle $ have the same frequency and 
direction of oscillation as the background mode $\Phi(\eta)$, then its effect is indistinguishable
from the oscillation of $\Phi(\eta)$. But it would change the amplitude of the oscillation and 
thus it can change the effective mass of the $\chi$ particle, which shifts the parameters of 
\Eqn{LameeqnX} from resonance band to non-resonance band and forces the shut-down of parametric 
resonance. This process is known as the restructuring of resonance.  In modified case
$\tilde g^2\over \lambda$ should look like
\begin{eqnarray}\label{phi-eta}
 {\tilde g^2\over \lambda}=  {g^2\over\lambda}{\Phi^2(\eta)+\langle \delta\phi^2 \rangle\over\tilde\Phi^2}
 = {g^2\over\lambda}\left(1-9.2{\langle\delta\phi^2\rangle\over\tilde\Phi^2}+{\langle\delta\phi^2\rangle\over\tilde\Phi^2}\right)
 = {g^2\over\lambda}\left(1-8.2{\langle \delta\phi^2 \rangle\over\tilde\Phi^2} \right)
\end{eqnarray}
For the derivation of the second step see the relationship between $\Phi(\eta)$ and $\tilde\Phi$ in \cite{Greene:1997fu}.
We can see that for the increase in $\langle \delta\phi^2 \rangle $ in the same direction of
$\Phi(\eta)$ oscillation, $\tilde\kappa^2$ remains unchanged and 
$\tilde g^2\over \lambda$ decreases with time. For $k=0$ mode from the stability-instability chart 
of \cite{Greene:1997fu} we can see that the $\tilde g^2\over \lambda$ should have a shift of 
${\cal O}(1)$ from ${g^2\over \lambda}=2$ to stop the resonance.

On the other hand if the leading mode of $\langle \delta\phi^2 \rangle $ has much higher 
frequency than the $\Phi(\eta)$ and comparable amplitude to it, this can go in negative
direction in a short time range. So that in a short time span the oscillating function 
of the \Eqn{LameeqnX} i.e. the $\cn(x,{1\over\sqrt{2}})$ will be modified and the parametric
resonance will be stopped. 

Among all the effects described above one or more than one of them can be responsible for
the end of preheating. It varies from model to model which particular process plays more 
important role than others. For our model $\lambda\phi^4$ it has been shown\cite{Greene:1997fu}
that the decrease in $\Phi(\eta)$ and thus the restructuring of resonance is the main cause for
the end of preheating. Although calculations in that work did not consider the presence of 
background $\chi$ field value. 

\section{Effect on large scales of CMB}\label{rev-large-scales}
Modes of the curvature perturbation which crossed horizon during initial period of inflation has
much larger wavelength than those modes which leaves horizon during  end of inflation. Minimum 
resolvable wavelength in CMB is around $e^{15}$ times smaller than the largest wavelength. So 
if inflation ends at 60 $e-$foldings   after the largest wavelength crossed the horizon, wavelengths
of interest during preheating is $e^{45}$ times smaller than the minimum resolvable wavelength in CMB. 
This huge hierarchy of scales ensures that even if a sizable amount of non-gaussian curvature 
perturbation is produced during preheating its effect on large scales CMB might be washed away. 
To predict how much imprint this small scale dynamics can leave on CMB the following method was 
developed\cite{Lyth:2005qk,Boubekeur:2005fj}. 

Let's say $H_i$ is the Hubble parameter at the end of inflation. Therefore amplitude of the power 
spectrum of the $\chi$ field's perturbation is $P_{\chi}={H_i^2}$. So, the spatial variance of
the $\chi$ field can be calculated as\cite{Lyth:2005du} 
\begin{eqnarray}
 \langle\chi(x)^2\rangle=\int {d^3k\over(2\pi)^3 }{P_{\chi}\over k^3}={P_{\chi}\over (2\pi)^2}\ln(q_{\rm max} L)
\end{eqnarray}
where $L$ is the length of the box size in which perturbation is defined and $q_{\rm max}$ is 
the frequency corresponds to $H_i$. From our above discussion it is evident that for our scales 
of interest $\ln(q_{\rm max} L)\approx 45$. We will use this spatial variance as $\sigma$ of a Gaussian
function to be used for smoothing out the curvature perturbation of small scales.
 \begin{eqnarray}
 \sigma^2(L) = \left(H_i\over 2\pi \right)^2 45
 \end{eqnarray}
 If $W$ is a Gaussian function with this sigma, then smoothed $\delta N$, say $\delta N_R$, is written as 
\begin{eqnarray}\label{gaussiansmooth1}
\delta N_R (\chi') = \int_{-\infty}^{\infty} \delta N(\chi) W(\chi'-\chi) d\chi .
\end{eqnarray}

We have two scalar fields in our system, inflaton $\phi$ and the secondary field
$\chi$. So we write down the curvature perturbation on large scales following \Eqn{zetaexpandx} as 
\begin{eqnarray}\label{zetaexpandx2}
\zeta(x)={ N^{\rm inf}_\phi}\delta\phi(x)+{ N^{\rm pre}_\phi}\delta\phi(x)+
{ N^{\rm pre}_\chi}\delta\chi(x)+{1\over2}{ N^{\rm
pre}_{\chi\chi}}\delta\chi^2(x)   .
\end{eqnarray}
Here two things are assumed, firstly the contribution of $\chi$ field during 
inflation is negligible and secondly inflaton $\phi$ does not produce any
second order curvature perturbation during preheating.
If we go to the Fourier space we get
\begin{eqnarray}
 \zeta_k=\underbrace{{ N^{\rm inf}_\phi}\delta\phi_k+{ N^{\rm
pre}_\phi}\delta\phi_k}_{\zeta_k^{\phi}}+
 \underbrace{{ N^{\rm pre}_\chi}\delta\chi_k+{1\over2}{ N^{\rm
pre}_{\chi\chi}}\int{d^3p\over(2\pi)^3}\delta\chi_p\delta\chi_{k-p}}_{\zeta_k^{
\chi}}.
\end{eqnarray}
We take one more assumption that is the contribution from perturbation of $\phi$ and $\chi$ are
uncorrelated i.e. $\langle\zeta_k^{\phi}\zeta_k^{\chi}\rangle=0 $ which allows us to write
the 2-point correlation as
\begin{eqnarray}
 \langle\zeta_{k_1}\zeta_{k_2}\rangle & = & \langle
\zeta^\phi_{k_1}\zeta^\phi_{k_2}\rangle + \langle
\zeta^\chi_{k_1}\zeta^\chi_{k_2}\rangle \nonumber\\
 & = &  N_\phi^2\langle\delta\phi_{k_1}\delta\phi_{k_2}\rangle+  (N^{\rm
pre}_\chi)^2
 \langle\delta\chi_{k_1}\delta\chi_{k_2}\rangle+ \nonumber \\
 & & {1\over4}\left(N^{\rm pre}_{\chi\chi}\right)^2\int{d^3p_1
d^3p_2\over(2\pi)^6}\langle\delta\chi_{p_1}\delta\chi_{k_1-p_1}\delta\chi_{p_2}
\delta\chi_{k_2-p_2}\rangle
\end{eqnarray}
and the 3-point correlation as
\begin{eqnarray}\label{three-pt-fn}
 \langle\zeta_{k_1}\zeta_{k_2}\zeta_{k_3}\rangle 
 & = & {1\over 2}  (N^{\rm
pre}_\chi)^2 N^{\rm pre}_{\chi\chi}\int{d^3p_1
\over(2\pi)^3}\{\langle\delta\chi_{p_1}\delta\chi_{k_1-p_1}\delta\chi_{k_2}
\delta\chi_{k_3}\rangle + {\rm permutations}\}
 + \nonumber \\
 & & {1\over8}\left(N^{\rm pre}_{\chi\chi}\right)^3\int{d^3p_1
d^3p_2d^3p_3\over(2\pi)^9}\langle\delta\chi_{p_1}\delta\chi_{k_1-p_1}\delta\chi_{p_2}
\delta\chi_{k_2-p_2}\chi_{p_3}\delta\chi_{k_3-p_3}\rangle
\end{eqnarray}
Defining $F_{\rm NL}$ as the ratio of bi-spectrum to the squared of power spectrum 
\cite{Komatsu:2002db,Bartolo:2004if,Liguori:2005rj} we get 
\begin{eqnarray}\label{fnl}
 F_{\rm NL}^{\rm local} & = & \left. {5\over 3}{\langle\zeta_{k_1}\zeta_{k_2}\zeta_{k_3}\rangle
 \over \langle\zeta_{k_1}\zeta_{k_2}\rangle^2+ {\rm permutations} }\right|_{k_1=k_2=k\gg k_3}\nonumber \\
 & = &  {5\over 6}(N^{\rm pre}_\chi)^2 N^{\rm pre}_{\chi\chi}{P_{\chi}^2\over P_{\zeta}^2}+ 
 {5\over 48}(N^{\rm pre}_{\chi\chi})^3{P_{\chi}^3\over P_{\zeta}^2}
\end{eqnarray}
In general local form non-gaussianity is defined to be the case where $\zeta$ can be 
written as $\zeta_g+{3\over 5} f_{\rm NL}\zeta_g^2$. This is not possible in case of 
our interest. But exactly like the standard local form non-gaussianity, this kind of non-gaussianity 
is also expected to show up in the squeezed limit of the bispectrum.


From the discussions presented at the beginning of this subsection we can say that here $\delta N_R$
will play the role of $\delta N^{\rm pre}$. So we can understand that depending on the form of
$\delta N_R$ either the first term or the second term in \Eqn{fnl} will play important role in the
value of $f_{\rm NL}$. But all these discussions are based upon one assumption that is expansion
of $\delta N_R$  can be terminated upto certain order. If it is not true then the situation become
more complex\cite{Suyama:2013dqa}. 

\section{Calculation of $\delta N$}\label{genform}
In the evolution of $H(t)$ three energy terms contribute, viz, potential energy, kinetic energy
and gradient energy. But upto the last stage of preheating we can neglect the contribution from
gradient energy. This assumption enables us to cast the Friedmann equation into the following form, 
\begin{eqnarray}\label{freedman-1}
\dot H(t)+3 H^2(t)\approx 8\pi V(t)\,\,\, ,
\end{eqnarray}
where dot means derivative with respect to $t$. In the above equation amplitude of $V(t)$ drops gradually with time. But if we
change the variables as, $ d\tau = dt/a(t)^2, {\cal H}=a^2 H, \Phi= a \phi$ and
$X= a\chi$, the amplitude of ${\cal V(\tau)}$ remains almost constant with $\tau$. 
Hence \Eqn{freedman-1} is rewritten as,   
\begin{eqnarray}\label{freedman}
{d{\cal H}(\tau)\over d\tau}+{\cal H}^2(\tau)\approx 8\pi {\cal V}(\tau) \,\,\, ,
\end{eqnarray}
where ${\cal V}$ becomes
\begin{eqnarray}\label{scaled-potential}
 {\cal V}= {\lambda\over4}\Phi^4+{g^2\over 2}\Phi^2 X^2.
\end{eqnarray} 

\begin{figure*}
\centerline{\epsfxsize=0.8\textwidth\epsfbox{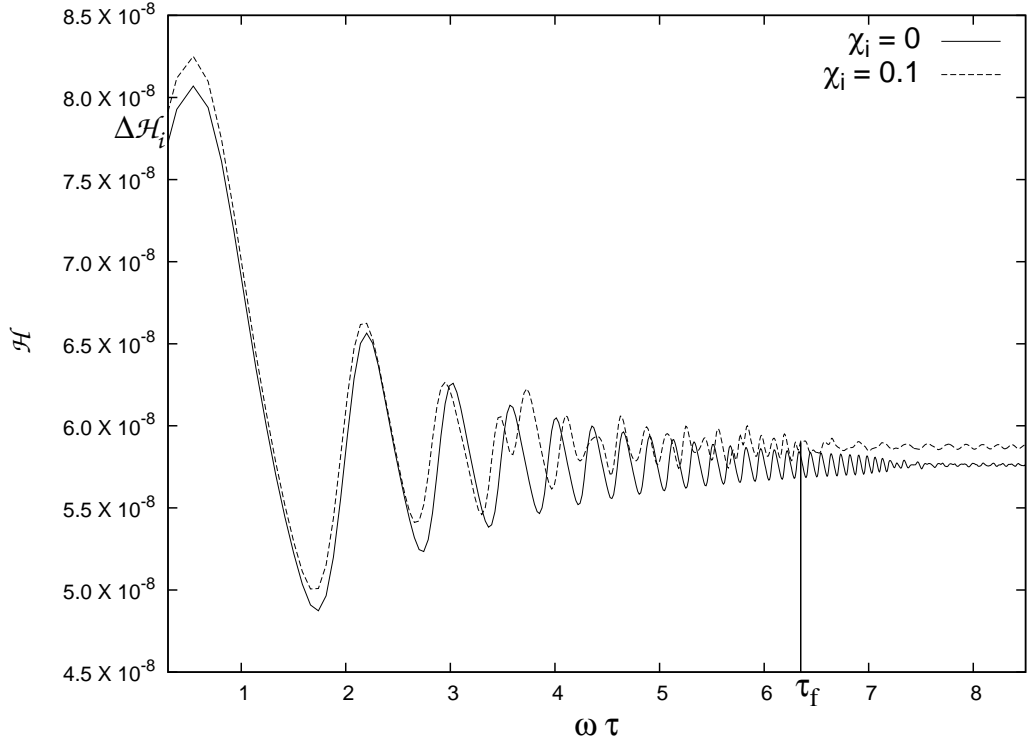}}
\caption{$\cal H$ trajectories oscillates with $\tau$ and unlike cases with fixed singular points
different trajectories with different initial $\cal H$ values cross each other. $\tau_f$ is the
time when the oscillation in the in $\cal H$ trajectory freezes, which means that preheating process
has been finished. Naturally $\tau_f$ varies for different trajectories. 
This plot is generated with lattice simulation for ${g^2\over \lambda}=2$. A very high value
of $\chi_i=0.1$ has been taken to magnify the difference in the $\cal H$ trajectories to a visible level.}
\label{trajectory-plot}
\end{figure*}

During the period
of preheating process, ${\cal V}$ is an oscillatory function. 
In the evolution of ${\cal H}$, we would have seen standard saddle-node
bifurcation feature from this equation if ${\cal V}$ was constant with $\tau$
That means in the
one-dimensional space of $\cal H$, there would have been a saddle point from which $\cal H$
would have move away while $\tau$ increases. There should be another stationary
point in this one-dimensional space which attracts the trajectories towards it. 
That point is termed as node. We
have checked that this saddle-node feature in $\cal H$ trajectories remains
even if $\cal V$ oscillates with constant amplitude (see \Fig{trajectory-plot}). 
Since initial values of $\cal H$ are always
positive, all the ${\cal H}(\tau)$ trajectories move towards node and away from
saddle.

Now we divide the total time span of the integration of \Eqn{deltaN} in two different
parts. First part is the period of exponential growth of the $X$ field which starts
from $\tau=0$ and grows upto $\tau=\tau_f$. Second period starts at $\tau=\tau_f$ and
continues upto $\tau=\tau_e$. The latter period behaves mostly like radiation dominated era.
So, the \Eqn{deltaN} can be re-casted into  
\begin{eqnarray}\label{limit-div}
 \delta N(\chi_i)=\int_{0}^{\tau_f}\left({\cal H}(\chi_i,\tau)-{\cal H}(0,\tau)\right)d\tau
 +\int_{\tau_f}^{\tau_e}{\cal H}(\chi_i,\tau)d\tau-\int_{\tau_f}^{\tau_e^0}{\cal H}(0,\tau)d\tau \,\,\, .
\end{eqnarray}
Here superscript ``0" indicates the value 
of these time limits for $\chi_i=0$ trajectory. Both the trajectories reach at Hubble value $H_e$ 
at $\tau_e$ and $\tau_e^0$ respectively. 

In this latter period, i.e., from $\tau_f$ to $\tau_e$, we assume the scale factor $a$ changes with 
physical time $t$ as $a(t)=\left(a_f^{1/\alpha}+ ct-ct_f\right)^\alpha$. If $w$
is the equation of state parameter then $\alpha={2\over3(1+w)}$ and $c$ is a constant whose value is given by
initial energy density.
This gives
\begin{eqnarray}
 \int_{\tau_f}^{\tau_e}{\cal H}d\tau=\int_{t_f}^{t_e} H dt= \int_{{\cal H}_f/a_f^2}^{H_e} {H\over\dot H} dH
 = -\alpha \log H_e +\alpha\log {\cal H}_f - 2\alpha\log a_f \,\,\, .
\end{eqnarray}
Putting this back in \Eqn{limit-div} we get
\begin{eqnarray}\label{limit-div-2}
 \delta N(\chi_i) & = & \int_{0}^{\tau_f}{\Delta\cal H}(\chi_i,\tau)d\tau
 +\alpha\log {{\cal H}_f(\chi_i)\over{\cal H}(0,\tau_f)} - 2\alpha\log {a_f\over a^0(\tau_f)}\,\,\, ,
\end{eqnarray}
where ${\Delta\cal H}(\chi_i,\tau) = {\cal H}(\chi_i,\tau)-{\cal H}(0,\tau)$.

As expected $H_e$ cancels out and $\delta N(\chi_i)$ remains independent of it.
Generally $w$ is expected to be ${1\over3}$ for radiation dominated era but lattice simulation
shows it is slightly less than that~\cite{Podolsky:2005bw}.
Therefore following that for our future purpose we will
choose $w=0.28$,{\it i.e}, $\alpha=0.52$.

Separation between $\cal H$ trajectories tends to decrease or increase with $\tau$.
This rate of decrease or increase in this difference is
characterized by the Lyapunov exponent ($\Lambda$) simply as,
\begin{eqnarray}\label{lyapunov-1}
\Delta{\cal H}(\tau)=\Delta{\cal
H}_{i} e^{\Lambda \tau}\,\,\, ,
\end{eqnarray}
where $\Delta{\cal H}(\tau)$ is the separation between the trajectories at a particular $\tau$, and 
$\Delta{\cal H}_i$ is the initial difference between two $\cal H$'s. These difference is introduced 
by their initial field values.

\subsection{Effect of node shift}
Separation of trajectories in \Eqn{lyapunov-1} would have been sufficient in the calculation of
\Eqn{limit-div-2} if the amplitude of $\cal V$ 
was constant with $\tau$ as well as with $\Delta{\cal H}_{i}$. But in reality, for different $\chi_{i}$, i.e., 
for different $\Delta{\cal H}_{i}$ preheating ends at different
values of $\tau$. The end values of $\cal H$ are also different 
as shown in \Fig{trajectory-plot}. So, the node point changes with initial conditions. 
Therefore the stationary points are movable. There is no definite 
prescription in the literature to answer how this change in node value modifies \Eqn{lyapunov-1}.
Here we take a quasi-static approach to see its effect. Node value in the \Eqn{lyapunov-1}
 is $\sqrt{8\pi{\cal V}}$ (say $A$). 
Amplitude and frequency of ${\cal V}$ vary with $\chi_i$. 
So change in $A$  at a particular $\tau$ can be written as
\begin{eqnarray}\label{A-expand}
 \Delta A= {\partial A\over\partial \chi_i} \delta\chi_i + {\partial A\over\partial \omega}
 {\partial\omega\over\partial\chi}\delta\chi_i\,\,\, ,
\end{eqnarray}
where $\omega$ is the frequency of oscillation of ${\cal V}$.

Solution of \Eqn{lyapunov-1} with constant $\cal V$ is 
\begin{eqnarray}\label{sol-wo-shift}
 {\cal H}(\tau)=A \left(\tanh\left(\tau A+\tanh^{-1}{{\cal H}_i\over A}\right)\right) \,\,\, .
\end{eqnarray}
If $A$ doesn't change with $\Delta{\cal H}_i$ it can be shown that, 
\begin{eqnarray}
 \Delta{\cal H}(\tau)= \Delta{\cal H}_i [B({\cal H}_i,A,\tau) A] \,\,\, ,
\end{eqnarray}
where $B({\cal H}_i,A,\tau)$ is a function of ${\cal H}_i,A$ and $\tau$, whose exact 
expression is shown in Appendix-\ref{nodeshiftcal}. To calculate Lyapunov 
exponent for constant $\cal V$ case we write \Eqn{freedman} in the form of difference equation as,
\begin{eqnarray}\label{difference_eqn_A}
 {\cal H}_{n+1}=\underbrace{{\cal H}_n- {\cal H}_n^2 + A^2}_{f_n}\,\,\, .
\end{eqnarray}
 Definition of Laypunov exponent is\cite{Arfken,Strogatz}
 \begin{eqnarray}\label{lyapunov_def}
  \Lambda={1\over n}\sum_i^n \log\left|\partial f_i\over\partial {\cal H}\right|
  \approx{1\over\tau}\int_0^{\tau}\log\left|\partial f\over\partial{\cal H} \right| d\tau' \,\,\, .
 \end{eqnarray}

If $A\ll 1$, which is true 
in our case, $\log[B({\cal H}_i,A,\tau) A$] matches quite well with the standard definition of Lyapunov 
exponent multiplied by $\tau$. So we can write 
\begin{eqnarray}
 B({\cal H}_i,A,\tau) A \approx e^{\Lambda\tau}
\end{eqnarray}

We assume the form of solution of \Eqn{sol-wo-shift} can describe the behaviour of ${\cal H}$ when $\cal V$ changes with time.
This is where the quasi-static approximation comes into play. With this assumption we 
write the difference in the $\cal H$ trajectories as, 
\begin{eqnarray}\label{sol-w-shift}
 \Delta{\cal H}(\tau)& = & {\cal H}(\tau,\tilde A,{\cal H}_i+\Delta {\cal H}_i )-{\cal H}(\tau,A,{\cal H}_i)\nonumber \\
                     & = & {{\cal H}(\tau,\tilde A,{\cal H}_i+\Delta {\cal H}_i )-{\cal H}(\tau,A,{\cal H}_i)\over 
                            B({\cal H}_i, A,\tau) A} e^{\Lambda\tau} \nonumber \\
                     & = & (\Delta {\cal H}_i + C (H_i,A,\tau) \Delta A ) e^{\Lambda\tau}  \,\,\, ,    
\end{eqnarray}
where $\tilde A = A + \Delta A$ and $C$ is a function whose value is of ${\cal O}(1)$ (see Appendix~\ref{nodeshiftcal}).
We will plug in this form of $\Delta{\cal H}(\tau)$ in \Eqn{limit-div-2} 
to calculate the value of $\delta N(\chi_i)$.
 
\subsection{Lyapunov Exponent}
So far we have dealt with constant $A$. Now we will vary $\cal V$ for calculation of $\Lambda$. So
the difference equations~\Eqn{difference_eqn_A} reads as,
\begin{eqnarray}
{\cal H}_{n+1}=\underbrace{{\cal H}_n- {\cal H}_n^2 + 8\pi {\cal V}_n}_{f_n}\,\,\, .
\end{eqnarray}
So, using the definition in~\Eqn{lyapunov_def} we write  

\begin{eqnarray}
 \int e^{\Lambda\tau} d\tau & =& \int \exp\left\{\left({1\over
\tau}\int_0^{\tau} \log \left|{\partial f(\tau')\over \partial {\cal H}}\right|d\tau'\right)\tau\right\} d\tau
 = \int \exp\left\{\int_0^{\tau} \left( \log \left|{\partial f(\tau')\over \partial {\cal H}}\right| - {\cal H}\right)d\tau'\right\}d\eta \nonumber \\
&=& \int \exp\left\{\int_0^{\tau} \left(-4\pi{\partial {\cal K} \over \partial {\cal H}} + 4\pi{\partial {\cal V} \over \partial {\cal H}} + ...\right) d\tau'\right\}d\eta \nonumber \\
&=& \int \left(1+{\int_0^{\tau} 
\left(1-4\pi{\partial {\cal K}\over \partial{\cal H}}+4\pi{\partial {\cal
V}\over \partial{\cal H}} \right)d\tau'} + ... \right) d\eta \,\,\, ,
\end{eqnarray}
where $d\eta={dt\over a(t)}$ and ${\cal K}=a^4K$, with $K$ being the kinetic
energy. 
In the last step we have taken $\partial {\cal K}\over \partial {\cal H}$ and $\partial {\cal V}\over
\partial {\cal H}$ to be small compared to one. We have also used \Eqn{freedman} and ${\cal H}^2= {8\pi\over 3}({\cal V +
K})$ relation.
Now we go to a different set of variables ${\cal P}$ and $\cal E$ to perform this integration. 
\begin{eqnarray}
 {\cal E}&=&{\cal V}+{\cal K}\; , \nonumber \\ {\cal P}&=&{\cal V}-{\cal K}\; .
\end{eqnarray}
In these variables the continuity equation reads as
\begin{eqnarray}
 {\partial{\cal E}\over\partial\tau}+3{\cal H P}-{\cal H E}=0\; .
\end{eqnarray}
Using these we write
\begin{eqnarray}
\int_0^{\tau}
\left(-4\pi{\partial {\cal K}\over \partial{\cal H}}+4\pi{\partial {\cal V}\over
\partial{\cal H}}\right)d\tau'= \int_0^{\tau}4\pi {3\over 4\pi({\cal E}-3{\cal P})}
{d{\cal P}\over d\tau'} d\tau' = - \log|{\cal E}_0-3{\cal P}|+\log|{\cal E}_0-3{\cal P}_0|\; .
\end{eqnarray}
In last step we assumed that ${\cal E}$ does not change too much with respect to 
the change in $\cal P$. 
Here ${\cal E}_0$ is the initial value of $\cal E$. 
The quantity $({\cal E}_0-3{\cal P})$ has a maximum amplitude 
of $4{\cal E}_0$ and ${\cal P}_0$ is equal to ${\cal E}_0$. 
Therefore expanding the above result upto second order in ${\cal P}$ we get,    
 \begin{eqnarray}
 &  {1\over 2} \left(1- {\left({\cal E}_0 - 3{\cal P}\right)^2\over 16{\cal E}_0^2}\right) + ... - \log 2 &=
  {15\over32} + {3 {\cal P}\over 16 {\cal E}_0}- {9 {\cal P}^2\over 32 {\cal E}_0^2} + ... -\log 2\nonumber\\
& & = {15\over32}- {3 {\cal K}\over 16 {\cal E}_0} + {3 {\cal V}\over 16 {\cal E}_0} - {9 {\cal K}^2\over 32 {\cal E}_0^2} + {9 {\cal K} {\cal V}\over 16 {\cal E}_0^2} - {9 {\cal V}^2\over 32 {\cal E}_0^2} + ... -\log 2\; .
 \end{eqnarray}
 So in-total we write 
\begin{eqnarray}\label{deltaNtotal}
 \int e^{\Lambda\tau} d\tau & = & \int \left( 1 + {15\over32}- {3 {\cal K}\over16 {\cal E}_0} + {3 {\cal V}\over16 {\cal E}_0} - {9 {\cal K}^2\over32 {\cal E}_0^2} + {9 {\cal K} {\cal V}\over 16 {\cal E}_0^2} - {9 {\cal V}^2\over 32 {\cal E}_0^2} + ... -\log 2 \right)d\eta\; . 
\end{eqnarray}
Here $\cal K$ can be taken as ${1\over2}\Phi'^2+{1\over2}X'^2$ for prolonged period of
inflaton oscillations.

For the calculation of third term in \Eqn{limit-div-2} we need to know the value of $a_f$ from
where we can take the assumption of nearly radiation domination as ${\cal H}$ as well as $a$ have 
oscillatory features, we can take the value of $a_f$ to be averaged value of $a$ at $\tau_f$. 
So we write,
\begin{eqnarray}
- 2\alpha\log {a_f\over a^0(\tau_f)} = -2\alpha\int_0^{\tau_f} \Delta{\cal H}_{\rm avg} e^{\Lambda\tau}\big|_{\rm avg}\; .
\end{eqnarray}

 As discussed earlier $\cal V$ is an oscillating even function of $\tau$. So, the second part 
 of \Eqn{A-expand} is an odd function of $\tau$ and the first part is even. Therefore in the integration of 
 the first part of \Eqn{limit-div-2} we can expect, for the large values of $\tau$, the effect 
 of the odd part of $\Delta A$ will be washed away. So, keeping only the even part and using integration by parts we write
 \begin{eqnarray}\label{deltaNhalf}
  \int_{0}^{\tau_f}{\Delta\cal H}(\chi_i,\tau)d\tau \approx \left[ \left(\Delta{\cal H}_i+ C {\partial A\over\partial\chi_i}\delta\chi_i
  \right) \int e^{\Lambda\tau}d\tau \right]_0^{\tau_f} \,\,\, .            
 \end{eqnarray}

 \section{$\delta N$ for particular parameter values}\label{dN-vs-gl}
Here in this section we will first assume some simple form $\Phi$ and $X$
motivated by their field dynamics. Then we will use them in $\cal K$ and $\cal V$
so that we can perform the integration of 
\Eqn{deltaNtotal}. Next we will calculate the maximum value 
achievable by $X$ for zero as well as non-zero $\chi_i$ cases. Using this we will
calculate $\Delta H_f$ and $\Delta H_i$ and we will plug them into \Eqn{deltaNhalf}.
These calculations will enable us to compute $\delta N_{\chi_i}$ using \Eqn{limit-div-2}.

\subsection{Functional forms of $\Phi$ and $X$}\label{assumptions}
The equations of motion for $\Phi$ and $X$ and their solutions have been 
described in section~\ref{rev-preheating}. 
Elliptic function $\cn(x)$ can be expanded in a series of cosine
functions. We will only consider the first term for our calculation, 
which is proportional
to $\cos{(0.8472 x)}$. The solution of the $k$-th mode of $X$ takes the
form $X_k=P(x)\exp(\mu_k x)$ where $P(x)$ is a
periodic function in $x$. Since we have already approximated $\cn(x)$ with
$\cos(0.8472 x)$ we can write down the Lame equation of $X$ in Mathieu equation
form. From there we find that $P(x)$ can approximated as $\cos(n x)$, where $n$
can be any real number. This $n$ gets different values for different ${g^2\over\lambda}$. 
So we write,
\begin{eqnarray}\label{X-form}
\Phi(\eta)&=& \tilde\Phi \cos(\omega\eta) \; ,\nonumber \\
X(\eta) &=& X_i \cos(n\omega\eta) e^{\mu\omega\eta}\; ,
\end{eqnarray}
where $\tilde\Phi$ is the initial value of $\Phi$ field and 
$X_i$ is the initial background value of $X$ field. If for certain
${g^2\over\lambda}$, background mode of $X$ field doesn't get amplified, then
this form of $X(\eta)$ might not hold. But in the cases where it does get
amplified, growth of the background mode overshadows the other non-leading modes. 
In \Eqn{X-form} $\omega= 0.8472 \sqrt{\lambda}\tilde\Phi$ and $\mu$ is the maximum
of the characteristic exponents $\mu(k)$
Integration limit $\eta_f$ is the time when $X$ reaches 
its maximum amplitude $X_{\rm max}$. 
This gives us
 \begin{eqnarray}
\eta_f= {\log{X_{\rm max}\over \chi_i}\over \mu \omega}\; .
\end{eqnarray}


Using these assumptions now we write 
\begin{eqnarray}
 {\cal V}={1\over4}\lambda\Phi^4\cos^4(\omega \eta)+{1\over 2}g^2\chi_i^2e^{2\mu\omega\eta}\Phi^2 \cos^2(\omega\eta)\cos^2(n \omega\eta)\; ,
\end{eqnarray}
and 
\begin{eqnarray}
 {\cal K} &=& {1\over2}\omega^2\Phi^2\sin^2(\omega\eta)+{1\over2} e^{2 \eta \mu \omega}\omega^2\mu^2\chi_i^2\cos(n\omega\eta)^2 \nonumber\\
 & - &  e^{2 \eta \mu \omega}\omega^2 n\mu\chi_i^2 \cos(n\eta \omega) \sin(n \eta\omega) + 
 {1\over2} e^{2\eta\mu \omega} n^2 \omega^2 \chi_i^2 \sin(n\omega\eta)^2\; .
\end{eqnarray}
We have deliberately written $X_i = \chi_i$ because $a=1$ at initial time and the form of
\Eqn{X-form} is suitable for the cases where background mode gets the maximum 
amplification.
Now we can plug in these forms in \Eqn{deltaNtotal} and do the integrations.

For the calculation of \Eqn{deltaNhalf} we need the initial value of $\Delta {\cal H}$
and the node shift value ${\partial A\over \partial \chi}\delta\chi$ at initial and final times.
Calculation of the initial value is straightforward and can be written as,
\begin{eqnarray}
 \Delta{\cal H}_i = 
 \sqrt{{8 \pi\over 3}\lambda \left({1\over 4} \tilde\Phi^4 + {g^2\over2\lambda} \chi_i^2  \tilde\Phi^2\right)} - 
  \sqrt{{8 \pi\over 3} {\lambda\over 4} \tilde\Phi^4 }\; .
\end{eqnarray}
The difference in the node values at the end of preheating between 
two trajectories with $\chi_i$ equals to zero and $\chi_i$ non-zero (say, $\Delta{\cal H}_f$)
is
\begin{eqnarray}\label{deltaHf}
{\partial A\over \partial \chi}\delta\chi = \Delta{\cal H}_f &=& 
 \sqrt{8 \pi\lambda \left({1\over 4} \Phi^4(\eta_f) \cos^4(\omega_{\Phi}\eta_f) + {g^2\over2\lambda} X_{\rm max}^2\Phi^2(\eta_f) \cos^2(\omega_{\Phi}\eta_f)\cos^2(n\omega\eta_f)\right)} \nonumber \\
 & - & \sqrt{8 \pi\lambda\left( {1\over 4} \Phi^4(\eta_f) \cos^4(\omega_{\Phi}\eta_f) +{g^2\over2\lambda} X_{\rm sat}^2 \Phi^2(\eta_f) \cos^2(\omega_{\Phi}\eta_f)\cos^2(n\omega\eta_f)\right) }\; .
\end{eqnarray}
Here $ X_{\rm sat}$ is the value of the field for which  $X$ reaches its maximum amplitude in 
zero $\chi_i$ case and $\omega_{\Phi}$ is the frequency of oscillation of $\Phi$ at the 
last stage of preheating.

In performing the integration of \Eqn{deltaNtotal} we will not considered the change in the 
frequency of $\Phi$ oscillation with time. This is because the change in $\omega$ mostly
happens at the final stage of preheating and contribution of that small range of $\tau$
is negligible in integration of \Eqn{deltaNtotal}.

\subsection{Calculation of $X_{\rm max}$ }\label{calculationofxmax}
The field $X(\eta)$ can be thought of an exponentially growing function up to $\eta_f$.
But precisely speaking this co-moving 
time does not actually signify the end of resonance of the background mode of $X$~\cite{Kofman:1997yn}.
Rather background mode reaches the saturation value ($X_{\rm sat}$) little earlier than $\eta_f$. 
And after the saturation of the background mode, the inhomogeneous modes 
continue to grow for a little while. In this section we will attempt to calculate
$X_{\rm max}$ which is the maximum amplitude of $X$ from some arguments of
energy conservation and will see how it varies with initial background values of $\chi$.

 We have seen in Section~\ref{rev-preheating} that the main effect which terminates the resonance of 
the background mode of $X_k$, is the restructuring of the resonance band. Let $\xi$ be the value
of $\tilde g^2\over \lambda$ for which resonance terminates. Therefore we see from \Eqn{phi-eta} that 
$\langle \delta\Phi^2\rangle$ has to increase up to the following value,
\begin{eqnarray}
 \langle \delta\Phi^2\rangle = \left(1-{\lambda\xi\over g^2}\right){\tilde\Phi^2\over 8.2}
\end{eqnarray}
 
For the problem in our hand we have to figure out how this $\langle \delta\Phi^2\rangle$ 
changes with the initial background field values of $\chi$. Production of $\chi$ particle from the
background classical field $\phi$ has been described as the scattering of $\chi$ particles with the
homogeneous condensate of $\phi$ particles \cite{Khlebnikov:1996mc,Khlebnikov:1996zt}. But here the
opposite thing needs to be studied, i.e., the scattering of $\phi$ particle with the homogeneous
condensate of $\chi$ particle. Addressing this problem is out of the scope of this paper. 

Therefore we give some logical arguments following the line of \cite{Kofman:1997yn} to get the dependence 
of $\langle X^2 \rangle$ with $\chi_i$ and thus the relation between $X_{\rm max}$ and $\chi_i$. 
First we will look at the $\chi_i=0$ case. 
We assume that $\langle\delta\Phi^2\rangle$ is proportional to $X^2$. It is expected 
if the production of $\langle\delta\Phi^2\rangle$ is mainly due to the rescattering, 
not the parametric resonance.
If $\langle \delta\Phi^2\rangle = \sigma X^2$ then we can write the 
saturation value of $X^2$ as 
\begin{eqnarray}
 X^2_{\rm sat}= \left(1-{\lambda\xi\over g^2}\right){\tilde\Phi^2\over 10.3 \sigma}\; . 
\end{eqnarray}

For example in ${g^2\over\lambda}=2$ case, the lattice result gives us the 
value of $X^2_{\rm sat}=0.39$ (see \Fig{Xsat-plot}). 
From the discussions in subsection~\ref{endofpreheating} it is evident that $\xi=1$.
Hence we can determine the value of $\sigma$ to be 0.05. Now we move on to
calculate the relation between the kinetic energies of these fields. Since
$\langle \delta\phi^2\rangle$ gets its kinetic energy from the 
kinetic energy of $X$ particle we can write

\begin{eqnarray}\label{kineticratio}
 {\kappa_{\Phi}^2\over 2}\langle \delta\Phi^2\rangle = \theta  {\kappa_X^2\over 2}\langle \delta X^2\rangle\; . 
\end{eqnarray}
Here $\kappa_\Phi$ and $\kappa_X$ is the leading mode in $\langle \delta\Phi^2\rangle$ and $\langle \delta X^2\rangle$
respectively and $\theta$ is the fraction which determines the amount of kinetic energy transfered to $\phi$ particle.
We know that the Lame equation \Eqn{LameeqnX} and \Eqn{Lameeqnphi} for $X_k$ and $\Phi_k$ respectively 
are just the same equation with different $g^2\over\lambda$ parameters. So from the stability-instability chart of
\cite{Greene:1997fu} we can figure out that $\kappa_X$ is much lower than
the $\kappa_\Phi$. We find $\kappa_\Phi^2\over \kappa_X^2$ is close to 16. 
Using the value of $\sigma$ we determine the value of $\theta=0.8$. 
So the way we have calculated $\theta$ for ${g^2\over\lambda }= 2$ case, we can 
determine its value for other $g^2\over\lambda$ cases.

Now we introduce nonzero value of $\chi_i$. One can think of it 
as the homogeneous condensate of particle and its kinetic energy will contribute to the available
kinetic energy for the production of $\Phi$ particles. So
we expect that the amplitude of $\langle \delta\Phi^2\rangle$ increases due to it. 
We have checked with 
lattice simulation \cite{Felder:2000hq} that number density of $\Phi$ particles increases 
substantially with the introduction of $\chi_i$. 
That means although the parametric resonance of the background mode and the low momentum modes of $X$ 
end earlier than that of the 
zero $\chi_i$ case, the extra growth of $\langle \delta\Phi^2\rangle$ will decay into $X$ 
particle and produce some more $\langle \delta X^2\rangle$. We write $X_{\rm max}^2$ as the combination
of $X_{\rm sat}^2$ and this extra $\langle \delta X^2\rangle$.

For the given potential of \Eqn{scaled-potential}, the scattering cross-section 
from $\Phi$ to $X$ in vacuum is~\cite{Kofman:1997yn}
\begin{eqnarray}
\sigma_{\rm scatt} \approx {g^2\over 16\pi\tilde\Phi^2}\; . 
\end{eqnarray}
But the actual rate of scattering will be multiplied by the number density of the $\Phi$ particle ($n^\Phi_k$)
since there is a huge number of $\Phi$ and $X$ particles in the medium.  
Extra kinetic energy due to nonzero $\chi_i$ is of the order of ${\mu^2\omega^2\chi_i^2\over 4} e^{2\mu\omega\eta_f^0}$
with $\eta_f^0$ is the comoving time when $X$ reaches its saturation value $X_{\rm sat}$ for 
$\chi_i=0$ case. So the extra $\langle \delta\Phi^2\rangle$ can be determined using \Eqn{kineticratio}. Extra 
$\langle \delta X^2\rangle$ will be $\sigma_{\rm scatt}n_{\Phi}$ times the extra $\langle \delta\Phi^2\rangle$.
Therefore we can write  
\begin{eqnarray}\label{xmax}
 X_{\rm max}^2 =  X_{\rm sat}^2 + \beta\chi_i^2\; , 
\end{eqnarray}
where $\beta = \theta\sigma_{\rm scatt} n_{\Phi}{\mu^2\over\kappa^2_{\Phi}} e^{2\mu\omega\eta_f^0}$.
The above expression is an approximate expression and derived without detailed
calculation of scattering between $\Phi$ particle and homogeneous condensate of $X$ particles.
Any detailed calculation in this direction might modify~\Eqn{xmax} and thus the shape of 
$\delta N (\chi_i)$.

Determination of $X_{\rm sat}$ and $\eta_f^0$ has been done using lattice simulation. It could have
been calculated analytically by considering back-reaction under some approximate analytical method
like Hartree approximation, as it has been taken in \cite{Bassett:1999cg, Zibin:2000uw}. But in that
case there was a possibility of overestimating these quantities, i.e. underestimating the effect of 
back-reaction. In \cite{Finelli:2000ya} $\eta_f^0$ has been calculated to be around  
$80\over\sqrt{\lambda}\tilde\Phi$ for ${g^2\over\lambda}=2$. But lattice simulation(see \Fig{latticeX}) indicates it to be
$59\over\sqrt{\lambda}\tilde\Phi$. Since $X$ is an exponentially growing function such over estimate
would lead to a high value of $\delta N$, as expected in \cite{Bassett:1999cg, Finelli:2000ya, Moghaddam:2014ksa}.
But lattice result of $\delta N$\cite{Bond:2009xx} shows us that in small $\chi_i$ region values of 
$\delta N$ are well within the observed limit.

\subsection{${g^2\over\lambda}=2$ case}
\begin{figure*}
\vspace{-0.2cm}
\centerline{\epsfxsize=\textwidth\epsfbox{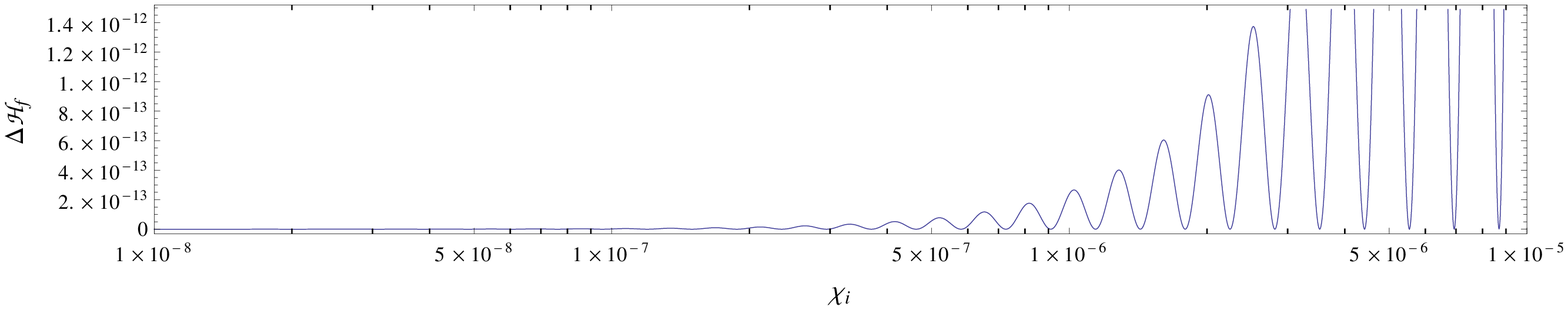}}
\centerline{\epsfxsize=\textwidth\epsfbox{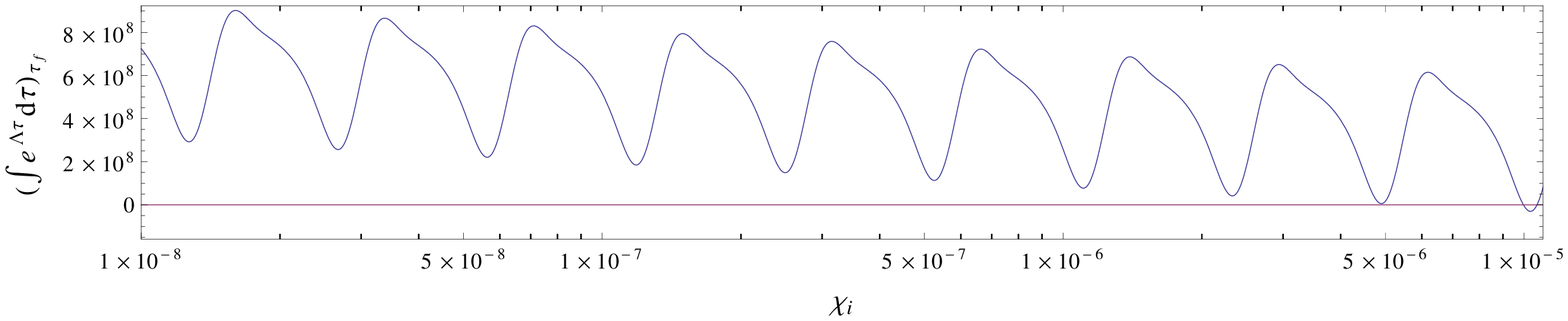}}
\centerline{\epsfxsize=\textwidth\epsfbox{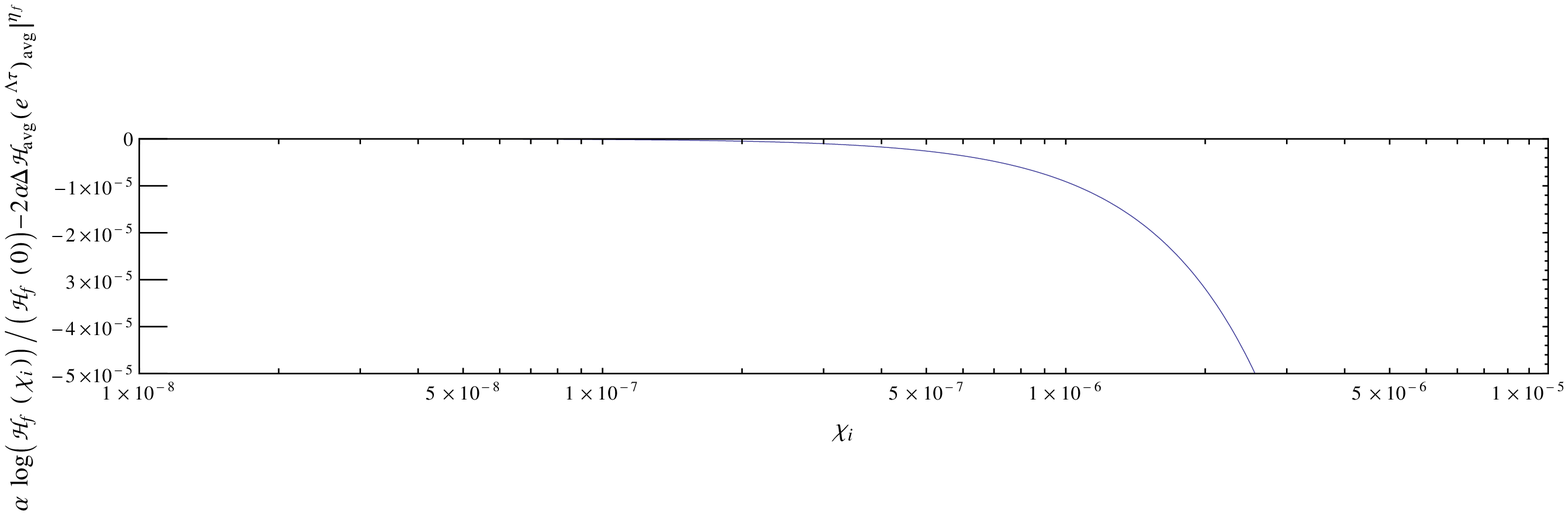}}
\centerline{\epsfxsize=\textwidth\epsfbox{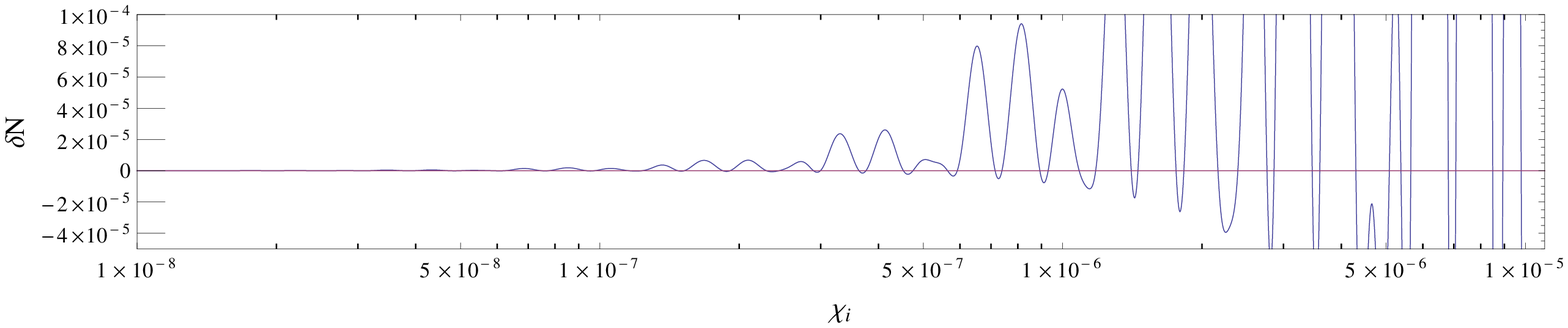}}
\caption{ This plot is drawn from \Eqn{limit-div-2} with $g^2\over\lambda$=2 and $\lambda=1\times
10^{-13}$. Since all others terms do not make substantial contribution these three quantities of the three 
upper panels make up the $\delta N$. First and second quantities gets multiplied and adds up to the third one
to produce total $\delta N$ in lowest panel.
Lattice result for $g^2\over\lambda$=2 has been shown in
\cite{Bond:2009xx}. }
\label{Fig:analyticsgl2}
\end{figure*}
Dynamics of preheating varies drastically for different values of ${g^2\over\lambda}$ parameter.
For ${g^2\over\lambda} =2$ case it is the $\kappa= 0$ mode which gets maximum amplification. 
Using the assumed form of $X(\eta)$, $\Phi(\eta)$ and other necessary
quantities as described in subsection~\ref{assumptions} and \ref{calculationofxmax} 
we evaluate \Eqn{deltaNtotal}.
For this choice of ${g^2\over\lambda}$, $X_{\rm sat}$ gets a value of 0.39 and $n$ becomes 1.
$\omega_{\Phi}$ comes out to be $\sim 3.3\omega$. Since at the end of preheating when resonance
stops we expect the frequency of $X$ remain equal to the driving frequency $3.3\omega$. This 
gives us the high frequency of $\Delta {\cal H}_f$. The ascending slope of $\Delta {\cal H}_f$
and thus $\delta N(\chi_i)$ comes from the relation between $X_{\rm max}$ and $X_{\rm sat}$, \Eqn{xmax}.
In the expression of $\beta$ of \Eqn{xmax} value of $n_{\Phi}$ has been taken to be 
as high as $10^{9}$ following the result of lattice simulation using \cite{Felder:2000hq} for 
non zero $\chi_i$. $\eta^0_f$ comes to be $59\over\sqrt{\lambda}\tilde\Phi$ (see Appendix \ref{deltaNcalculation}).

The first term in~\Eqn{limit-div-2} is calculated using~\Eqn{deltaNhalf}. In the combination
$\Delta {\cal H}_i + C \Delta {\cal H}_f$, contribution of $\Delta {\cal H}_i$ is negligible
compared to the latter term. Variation of $\Delta {\cal H}_f$ has been shown in first panel
of \Fig{Fig:analyticsgl2}.

Different quantities calculated in this process are shown in \Fig{Fig:analyticsgl2}. 
$\int_{0}^{\tau_f^0}{\cal H}(0,\tau)d\tau $
does not impart significant contribution in the $\delta N$ and 
integration of Lyapunov exponent is
also not sensitive to the last stage of the preheating. 
$\Delta{\cal H}_i$ is always negligible compared
to $\Delta{\cal H}_f$. There for only these three terms shown 
in this figure contribute. We do not show the total expression for 
$\delta N(\chi_i)$ to avoid complexity. But approximate expression
will be shown later. 

One can see the plot contains all the features shown in the lattice result 
in \cite{Bond:2009xx}. But the dependence of $X_{\rm max}$ with $\chi_i$ 
is an issue needs to addressed in much details. How can $\chi_i$  influence the increase 
in total $\chi$ particle production even after the end of 
parametric resonance has to be addressed to get the
right slope for the $\delta N$ with $\chi$. 
From our estimation o \Eqn{xmax} it comes out to be little higher 
than the slope in the lattice result of \cite{Bond:2009xx}. 
This is one of the main findings of our paper, that
the curvature perturbation depends not only on the parametric resonance 
but also on the finer dynamics of
preheating like rescattering and exact process responsible for the end of preheating.

\subsection{${g^2\over\lambda}=200$ case} 
\begin{figure*}
\vspace{-0.2cm}
\centerline{\epsfxsize=0.7\textwidth\epsfbox{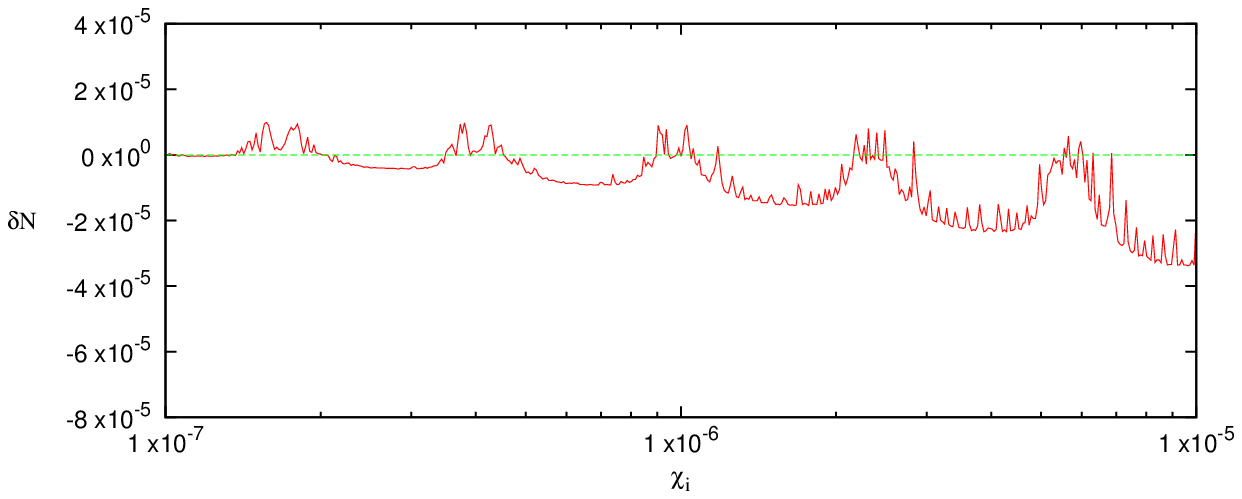}}
\centerline{\epsfxsize=0.7\textwidth\epsfbox{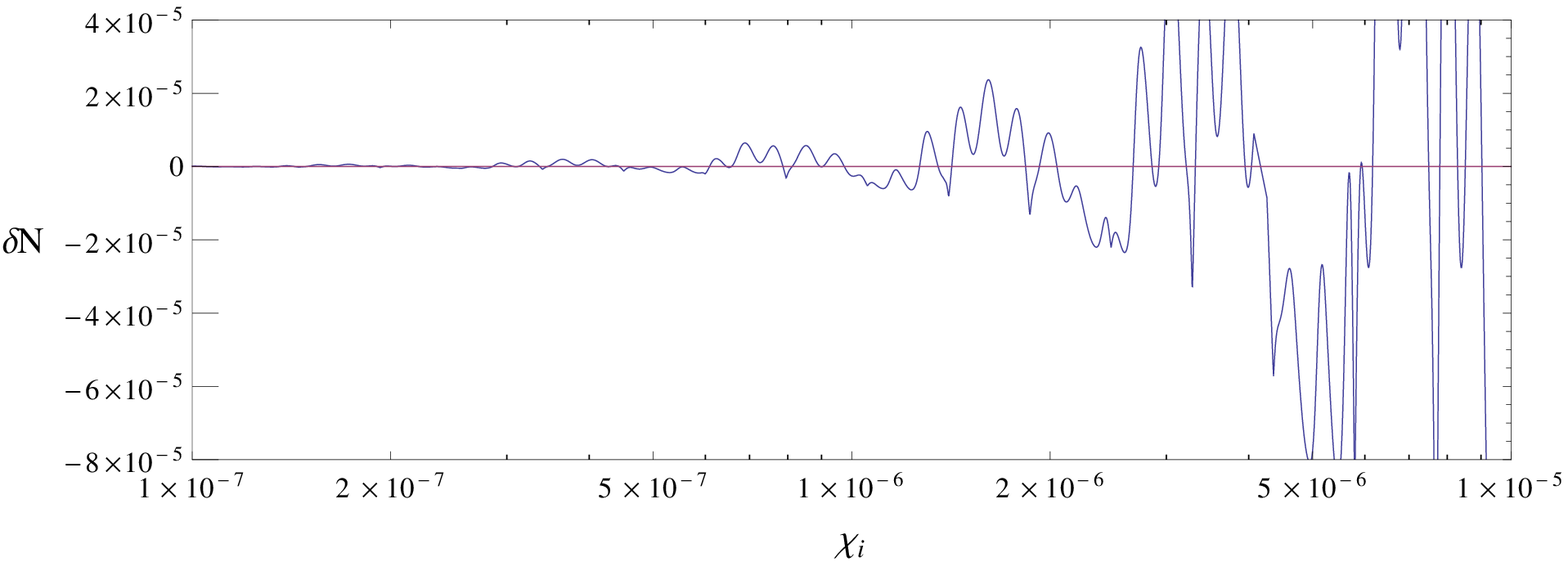}}
\caption{The upper panel shows lattice simulation of  $\delta N(\chi_i)$ for $g^2\over\lambda$=200. The lower panel shows the analytic
result following from \Eqn{limit-div-2} for this parameter choice.}
\label{Fig:plot-gl200}
\end{figure*}
To double check if our formalism is right or wrong we do the lattice simulation
for the same model with $g^2\over \lambda$=200 in
\Fig{Fig:plot-gl200}\footnote{Lattice simulation is done using publicly
available LatticeEasy code\cite{Felder:2000hq}. The default code cannot give
energy conservation accuracy(ECA) less than $10^{-3}$. We have achieved an ECA
of order $10^{-7}$ reducing the time step and using double precession. The above
plot is a result of simulations with 1000 different initial $\chi$ values on
$32^3$ lattice. Length of the lattice has been set at $20/H_e$, where 
$H_e$ is the value of Hubble parameter at the end of inflation}. 
Such a high value of $g^2\over\lambda$ is irrelevant for
the production of non-gaussianity in large scale. It makes $\chi$ field
heavier than Hubble parameter during inflation and thus no super-horizon perturbation
of $\chi$ gets produced. 
So applying $\delta N$ formulation during preheating era becomes impossible.

The dynamics of $\chi_i$ is totally different in this case. 
For this value of $g^2\over\lambda$ the 
$\kappa=0$ mode doesn't get amplified, but the higher $\kappa$ modes gets. 
Still they have quite small value of $\mu$. 
So in the \Fig{Xsat-plot} we can see that $\langle X^2\rangle$ initially grows slowly. The moment 
$\tilde g^2\over\lambda$ reduces to 180 the $\kappa=0$ mode starts increasing. At ${\tilde g^2\over\lambda}=178$
 it gets its highest value $\mu=0.23$. So we repeat the earlier formulation with this value of $\mu$ keep the 
 assumed form of $X(\eta)$ as \Eqn{X-form}. 
There are few more things which gets modified in this ${g^2\over\lambda}=200$ case than the previous one. Firstly
due to the change in $\xi$ and $g^2\over\lambda$ the value of $\langle X^2\rangle_{\rm sat}$ gets modified.
It comes down to $0.01$ and $\eta^0_f$ comes to be $48\over\sqrt{\lambda}\tilde\Phi$. $\omega_{\Phi}$ is 
$3 \omega$. One can numerically find out that the value of $n$ can be taken to be 7. $n_{\Phi}$ has been taken
to be of the order of $10^{6}$.

Unlike the ${g^2\over\lambda}=2$ case, at the end of resonance $X$ field gets a large non-oscillatory inhomogeneous
$\langle\delta X^2\rangle$, for both $\chi_i=0$ and nonzero case. So in $\Delta {\cal H}_f$ there is non-oscillatory
part. This part magnifies the effect of Lyapunov exponent in the plot of $\delta N(\chi)$ of \Fig{Fig:plot-gl200}.
We see $\delta N(\chi_i)$  matches well with the lattice result. As the earlier case exact match 
can only be possible if the exact dynamics of rescattering and backreaction is understood.

So in short from the above discussions we understand that the spiky patterns are the manifestation of the increase 
in the frequency of the homogeneous mode of $\Phi$ at the final period of preheating. As the frequency becomes
higher, the number of spikes increases. We also understand that the increase in the amplitude of $\delta N$ 
with $\chi_i$ comes from the fact that the saturation value of $X$ ($X_{\rm max}$) increases with the increase
in $\chi_i$. The dependence of $X_{\rm max}$ on $\chi_i$ determines how $\delta N$ would increase with $\chi_i$. 
From some logical arguments developed following the line of \cite{Kofman:1997yn}, we got 
$X_{\rm max} = X_{\rm sat} + \beta\chi_i^2$.
But to exactly determine the dependency of $X_{\rm max}$ on $\chi_i$, one needs to study the rescattering of
$\phi$ particles with the homogeneous condensate of $\chi$ particles. 
Since we have taken a $\chi_i^2$ dependence of $X_{\rm max}$, we have got a suppressed value of $\delta N$ 
for the lower values of $\chi_i$ and elevated values $\delta N$ for larger $\chi_i$.
From the lattice simulation of \cite{Bond:2009xx} it seems that $X_{\rm max}$ should depend on some lower power of $\chi_i$
than $\chi_i^2$ because in that case spikes in the lower $\chi_i$ region will be more visible.
\section{Non-gaussianity}\label{smoothing}
In the second part of our paper we move on to calculate the effect of curvature perturbations
on CMB, generated in preheating era. 
The exact functional form of the quantities shown in \Fig{Fig:analyticsgl2} is too much
complicated to analytically predict the value of any observable quantity. 
So we pick out a few terms from the total expressions of \Eqn{limit-div-2}
which are most dominating in yielding the shape of $\delta N$.
After calculations those terms give the following simplified form of $\delta N$,

\begin{eqnarray}\label{deltaNgl2}
 \delta N(\chi_i) & \approx & {\sqrt{8\pi} \over 2}C \left[{
 {g^2\over\lambda} \beta \Phi(\eta_f) \sqrt{\lambda (\Phi(\eta_f)^2 +
     2 {g^2\over\lambda} X_{\rm sat}^2)}\over
 2 \left(\Phi(\eta_f)^2 + 2 {g^2\over\lambda} X_{\rm sat}^2\right)^2}\chi_i^2\right]
 \nonumber\\
 & & \times \left({\left({47\over32} - \log 2\right)\over \mu\omega }\log{X_{\rm sat}\over\chi_i} 
 - {135\over 1024} {g^2\over \lambda}{X_{\rm sat}^2\over \tilde\Phi^2}
 {\mu \cos\left(2\log{X_{\rm sat}\over\chi_i}\right)+\sin\left(2\log{X_{\rm sat}\over\chi_i}\right)
 \over (1 + \mu^2) \omega}\right.\nonumber\\
 & & -\left. {27\over 256}{g^4\over\lambda^2}{X_{\rm sat}^4 \over \tilde\Phi^4}{2\mu\cos\left(2 \log{X_{\rm sat}\over\chi_i}\right)+
 \sin\left(2 \log{X_{\rm sat}\over\chi_i}\right)\over(1+4\mu^2)\omega}\right) \nonumber \\
 & & -\left[{ {g^2\over\lambda} \alpha \beta \left(-8 + 3 C \sqrt{2 \pi}
        \sqrt{\lambda \Phi(\eta_f)^2 (\Phi(\eta_f)^2 + 2 {g^2\over\lambda} X_{\rm sat}^2)}
        \left[\int e^{\Lambda\tau} d\tau\big|_{\eta_f}\right]_{\rm avg}\right)\over
 4 (\Phi(\eta_f)^2 + 2 {g^2\over\lambda} X_{\rm sat}^2)}\chi_i^2 \right]. 
\end{eqnarray}
This is the main result of our paper. In principle one can calculate the non-gaussianity
parameters using this expression of $\delta N$. There are two methods for this purpose 
in literature. First one Fourier decomposes $\delta N(\chi_i)$ in terms of variance and then
calculate the three point correlation function~\cite{Suyama:2013dqa}. We have failed to use
it for the $\delta N(\chi_i) $ of \Eqn{deltaNgl2}. So we follow the second method which has
been described in section~\ref{rev-large-scales}.

We smooth out \Eqn{deltaNgl2} on the large scales by the gaussian window function
as \Eqn{gaussiansmooth1}.
\begin{eqnarray}\label{gaussiansmooth}
 \delta N_R (\chi') = \int_{- \infty}^{\infty} \delta N(\chi)
W_{\chi}(\chi'-\chi) d\chi= 2 \int_{0}^{\infty} \delta N(\chi)
W_{\chi}(\chi'-\chi) d\chi\; .
\end{eqnarray}
The last step follows from the fact that $\delta N(\chi_i)$ depends only on the
absolute value of $\chi_i$. So, if we change the variable $\chi_i$ to $y= \log\chi_i$ then
\Eqn{gaussiansmooth} gets the following form,
\begin{eqnarray}\label{gaussianlogsmooth}
 \delta N_{R}(\chi')= 2 \int_{-\infty}^{\infty} \delta
N(y)\underbrace{{ e^{y}\over \sqrt{2\pi}\sigma}\exp\left(-{(e^y-\chi')^2\over \sigma^2}\right)}_{W_y}dy\; .
\end{eqnarray}
For computing the integration of \Eqn{gaussianlogsmooth}  analytically, we use the following method.
We expand $W_y$ in terms of a gaussian function $G_y$ using Edgeworth expansion~\cite{Bernardeau:1994aq}
\begin{eqnarray}\label{Edgeworth}
 W_y(y,\chi')=G_y(y,y_0)\left[1-K_1{y\over\gamma}+{K_2\over2\gamma^2}\left({y^2\over\gamma^2}-1\right)-{K_3\over
6\gamma^3}\left({y^3\over\gamma^3}-3{y\over\gamma}\right)+ ...\right]\; ,
\end{eqnarray}
where
\begin{eqnarray}
 G_y(y,y_0)={1\over\sqrt{2\pi}\gamma}\exp-\left({y-y_0\over\sqrt{2}\gamma}\right)^2 \, \, .
\end{eqnarray}
Here $K_i, y_0$ and $\gamma $ are functions of $\sigma$ and $\chi'$(see
Appendix \ref{Edgeworthexpansion}). $y_0$ is the value of $y$ for which $W_y$ becomes maximum.
So, $y_0$ is not necessarily the mean value of the distribution.
The coefficient $K_1$ takes care of the shift in the mean. Similarly $K_2$ and $K_3$
take care of the change in variance and skewness. For the values of model parameters we have taken,
$\sigma$ gets a value of $8.38\times 10^{-8}$. This expansion fits quite well for
$\chi'> 5\times 10^{-8}$. So whatever result we would obtain after integrating
\Eqn{gaussianlogsmooth} is valid for such a choice of $\chi'$.

After changing the variable to $y$ we get four types of terms. The contribution of the terms $e^{2y}\sin ny$ 
and $e^{2y}y\sin ny$ are proportional to $e^{2-{n^2\over 2}+2y_0}\sin[{n(2+y_0)}]$ 
after performing the interaction~\Eqn{gaussianlogsmooth}. Thus for large values of $\chi_i$
when $y_0$ is quite small, we can expect that the low frequency modulation contributes in the
smoothed out $\delta N_R$. This can be seen in the numerical integration performed 
in~\cite{Bond:2009xx}. We are not interested in such high values of $\chi_i$.
Therefore we are interested only in these two types of terms which are proportional to
$e^{2y}$ and $e^{2y}y$.

After doing the integration of \Eqn{gaussianlogsmooth} using \Eqn{deltaNgl2} we find $\delta N_R(\chi')$ takes the following form,
\begin{eqnarray}\label{deltaNR}
 \delta N_R(\chi')\approx e^{2(y_0 + \gamma^2)} \sqrt{\gamma} \left[C_1 + C_2(y_0 + 2\gamma^2)\right]\; , 
\end{eqnarray}
where the exact expressions of $C_1$ and $C_2$ are shown in Appendix~\ref{Edgeworthexpansion}. For more 
accurate determination of $\delta N_R$ higher order terms like $K_3$ or $K_4$ can be taken into 
consideration. But the form of \Eqn{deltaNR} remains same. Expanding \Eqn{deltaNR} in a converging 
series of $\chi'$ as \Eqn{zetaexpandx} or \Eqn{zetaexpandx2} is impossible for the large range of $\chi'$.
The reason behind this is that $\gamma, y_0$ and $C_1, C_2$ can not be expanded in converging series of $\chi'$. 
Still we find that the dependence of $\delta N_R$ on $\chi'$ is mostly proportional to $|\chi'|$. We
don't know what kind of signature it would produce in non-gaussianity parameters. Still to have an
idea of the order of $f_{\rm NL}$ we write approximately \Eqn{deltaNR} in a small range of 
$7\times 10^{-8}<\chi'<3\times 10^{-7}$ as
\begin{eqnarray}\label{smootheddeltaN}
\delta N_R(\chi')\approx  10^{9} {\chi'}^2 \; .
\end{eqnarray}
Using the second term of \Eqn{fnl} we get $F_{\rm NL}^{\rm local}\sim {\cal O}(1)$. We
have used $P_{\chi}\sim 2\times 10^{-15}$ and $P_{\zeta}\approx 2.4\times 10^{-9}$.

But as discussed earlier the value of $F_{\rm NL}$ can vary depending on the range of $\chi'$. So,
this result does not in any sense mean to be taken as prediction of $F_{\rm NL}$. 

%


\section{Conclusion}\label{conclusion}
In this work we have developed a technique for calculating curvature perturbation in the
era of preheating. During this era, different disconnected Hubble patches evolve
differently. We have used Lyapunov theorem to calculate the separation between
different evolution trajectories of Hubble parameters. This method has enabled us
to use $\delta N$ formulation and calculate a functional form of curvature perturbation
analytically. In this way we have been able to establish a one-to-one connection between
the finer dynamics of preheating and the super-horizon curvature perturbation.
We have shown the dependency of saturation value of the secondary field, $\chi$ with
its initial background value plays a crucial role in the expression of curvature
perturbation or $\delta N$.

The parametric resonance of $\chi$ ends via different dynamics for different models
of potentials. In some cases rescattering produces enough amount of inflaton
particles and forces the parametric resonance to shut down. In other models restructuring
of resonance might play the role. But availability of background field value of $\chi$
can act as a homogeneous condensate of $\chi$ particles and might effect the
process of rescattering. Similarly, change in inflaton particle production can alter
the production of $\chi$ particles just after the end of the parametric resonance.
These dynamics changes the saturation value of $\chi$ ($X_{\rm max}$) for
its different initial background values ($\chi_i$). Detailed calculation of
how $X_{\rm max}$ depends on $\chi_i$ has not been explored in previous literatures.
We have shown that this relation is the most important factor for the determination
of $\delta N(\chi_i)$.

Here we have calculated a relation between $X_{\rm max}$ and $\chi_i$ from some
simple logical arguments. Using that we have presented a simplified and approximated
form of $\delta N(\chi_i)$. This form matches quite well with the shape of $\delta N(\chi_i)$
computed using lattice simulation~\cite{Bond:2009xx}. To check whether our formulation is
correct or not, we have performed lattice simulation for another  set of 
parameter values and we have observed that it matches reasonably
well with the analytical calculation. The lattice results indicate that the relation between
$X_{\rm max}$ and $\chi_i$ needs to be modified to have an exact match. So future works are
required in this direction.

The second problem we have attempted to solve is how the curvature perturbation $\delta N$
produced in era of preheating can show up in large scales of CMB. From the functional form
of $\delta N(\chi_i)$ we see that it depends on $\log\chi_i$ that means as expected earlier
by various authors $\delta N$ would depend only on the absolute value of $\chi_i$.
We smooth out the $\delta N(\chi_i)$ with suitable Gaussian window function and we have given
an analytical form of curvature perturbation smoothed out on large scales ($\delta N_{R}$).
This $\delta N_{R}$ cannot be expanded in a converging series of $\chi$ for the entire range.
Therefore as expected by earlier authors to approximate $\delta N_{R}$ as a quadratic function
of $\chi$ for any arbitrary values of $\chi$. Although we have taken an approximate quadratic
form of $\delta N_{R}$ to have an estimation of the order of local form
non-gaussianity parameter $F_{\rm NL}$,
this estimation might not hold for any arbitrary values of $\chi_i$. So, we restrict ourselves
from predicting any particular value of non-gaussianity parameter $F_{\rm NL}$. Rather we pose
it as a problem to be addressed for the cases where $\delta N_{R}$ depends on $\log\chi$.

Therefore in total we have built a formulation for calculating curvature perturbation
from the dynamics of preheating and we have
methodically shown how some specific quantity of preheating
dynamics effects different features of $\delta N$. We have taken the massless
preheating model just to have a comparison with the available lattice simulation in
the literature. But, in general, the procedure developed here for calculating
$\delta N(\chi)$ can be applicable to any other potentials. Even for this potential one
can incorporate the contribution of more finer dynamics like the oscillation of inhomogeneous
modes of inflaton and see its effect on $\delta N(\chi)$. Therefore we hope that our study would
serve as a good platform for accurate determination of $f_{\rm NL}$ from preheating in future. 
\section*{Appendix}
\appendix
\section{Calculation for node shift dependence of $\Delta{\cal H}$}\label{nodeshiftcal}
Using \Eqn{sol-wo-shift}  we expand the difference between two solutions of $\cal H$
 with different initial condition
with respect to their initial difference $\Delta{\cal H}_i$ and take up to the first order term.
\begin{eqnarray}
\Delta{\cal H}(\tau)= {\cal H}(\tau,{\cal H}_i+\Delta{\cal H}_i)-{\cal H}(\tau,\Delta{\cal H}_i)= 
{A^2\over (A \cosh(A\tau) + {\cal H}_i \sinh(A\tau))^2}\Delta{\cal H}_i
\end{eqnarray}
So 
\begin{eqnarray}
 B={A\over (A \cosh(A\tau) + {\cal H}_i \sinh(A\tau))^2}
\end{eqnarray}
Now by introducing the change in $A$ we come up to \Eqn{sol-w-shift} where the function $C$ is
\begin{eqnarray}
 C({\cal H}_i,\tau, A)= {1\over 4 A^2} e^{
 2 A \tau} \left({\cal H}_i + e^{-2 A\tau} {\cal H}_i + 
   A- A e^{-2 A \tau} \right) ({\cal H}_i - e^{-2 A \tau}{\cal H}_i + 
   A + A e^{-2 A \tau } ) 
\end{eqnarray}
If write $A= 8 \pi {\cal V}$ and $\tau$ at the end of the preheating is of the order of ${\cal O}(10^2)\eta_f$ and
${\cal H}_i= {\lambda\over 4}\tilde\Phi^4$ we get $C\approx {\cal O}(1)$.

\section{Determination of the necessary quantities}\label{deltaNcalculation}
\begin{figure}\label{latticeX}
 \centerline{\epsfxsize=0.5\textwidth\epsfbox{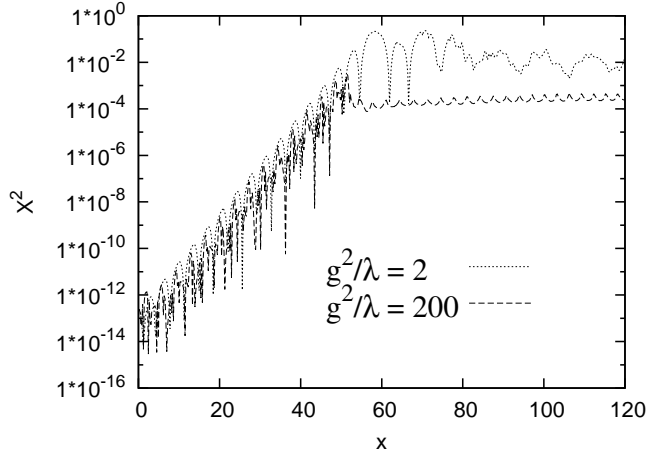}}
\caption{Growth of the field $ X^2 $ for different values of $g^2\over \lambda$. Here $x=\sqrt{\lambda}\tilde\Phi\eta$ 
as described in section \ref{rev-preheating}}
\label{Xsat-plot}
\end{figure}

We determine two quantities for the evaluation of \Eqn{deltaHf} and \Eqn{xmax} from 
lattice simulation results using \cite{Felder:2000hq}. First one is the co-moving 
time $\eta^0_f$ when the $X$ reaches its saturation value for $\chi_i=0$ case, and 
second one is the value of $X$ field at saturation
level called $X_{\rm sat}$. In the \Fig{latticeX} the growth of the $X$ field with 
$\sqrt{\lambda}\tilde\Phi\eta$ has been shown for different values
of $g^2\over\lambda$. For both the case $\chi_i$ has been taken to be zero. 
The maximum value reached by $X$ is recorded as $X_{\rm sat}$, and corresponding 
co-moving time as $\eta^0_f$. $X_{\max}$ in the \Eqn{xmax} is this maximum value of $X$
for nonzero $\chi_i$ cases. From the plots we get $X^2_{\rm sat}$ to be 0.39 and
$\eta^0_f={59\over \sqrt{\lambda}\tilde\Phi}$ for ${g^2\over\lambda}=2$. For 
${g^2\over\lambda}=200$, $X^2_{\rm sat}$ turns out to be and $10^{-4}$ and $\eta^0_f$ to be
${48\over \sqrt{\lambda}\tilde\Phi}$.

\section{Expansion of $W_y$}\label{Edgeworthexpansion}
\begin{figure*}
\centerline{\epsfxsize=0.5\textwidth\epsfbox{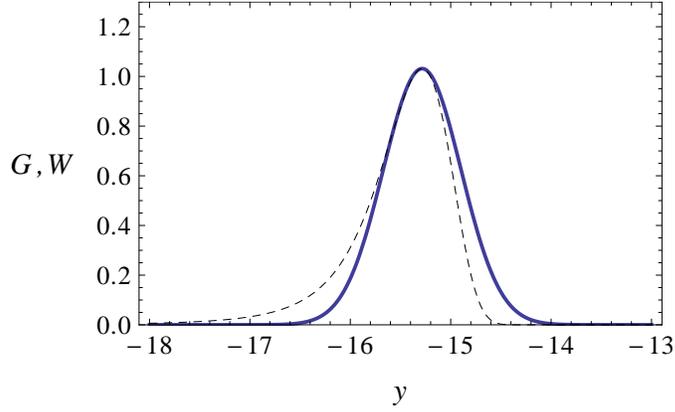}}
\caption{Probability distribution functions ($G$ and $W$) for $\log(x)=y$ with $\sigma=8.38\times10^{-8}$ and $\chi'=2\times10^{-7}$. $G_y$ has been shown with solid line and $W_y$ by dashed line. This indicates that slightly non-gaussian $W_y$ can be expanded in terms of fully gaussian $G_y$.}
\label{pdf-plot}
\end{figure*}
Edgeworth expansion is an expansion of slightly non-gaussian probability
distribution
function in a polynomial series multiplied by a gaussian function as shown in
\Eqn{Edgeworth}. First thing one need to do is to fix a suitable gaussian
function so that the error can be minimized. We choose the mean value ($y_0$) of the gaussian
function $G_y$ to be that value of $y$ for which $W_y$ reaches maximum. So, $y_0$ turns
out to be
\begin{eqnarray}
 y_0= \log\left[{1\over2} \left(\chi' + \sqrt{4 \sigma^2 + \chi'^2}\right)\right] .
\end{eqnarray}
Second thing to be fixed is the variance $\gamma$. To fix this we choose the
height of
the $G_y$ and the $W_y$ to be equal, which gives
\begin{eqnarray}
 \gamma^{-1} = {1\over 2\sigma }\left(\chi'+\sqrt{4\sigma^2+\chi'^2}\right)
\exp{\left[\left(-\chi'+{1\over2}\left(\chi'+\sqrt{4\sigma^2+\chi'^2}\right)\right)^2\over
2\sigma^2\right]}
\end{eqnarray}
It can be an interesting matter of study that which choice of $\gamma$ and
$y_0$ can give best fit up to a certain order of expansion, and how
does the value of $f_{\rm NL}$ depends on this choice. For this present paper we avoid this discussion. But just for an illustration
we plot the $W_y$ and the $G_y$ in \Fig{pdf-plot} with some realistic values of
$\sigma$ and $\chi'$. Now we move on to calculate $K_i$s. $K_1$ is defined as
\begin{eqnarray}
 K_1 &=& \int_{-\infty}^{\infty}\left({y-y_0 \over \gamma}\right)
 W_y=\int_{0}^{\infty}\left({\log x - y_0 \over \gamma}\right)
 {1\over \sqrt{2\pi}\sigma}e^{-{(\chi-\chi')^2\over2\sigma^2}}\nonumber\\
&= & \frac{1}{4 \sigma \gamma}e^{-\frac{\chi'^2}{2 \sigma ^2}}
\left\{-e^{\frac{\chi'^2}{2 \sigma ^2}} \sigma 
\left( \gamma_E + 2y_0 + \log 2 + {\rm {\rm Erf}}\left[\frac{\chi'}{\sqrt{2} \sigma }\right]
(\gamma_E+ 2y_0 + \log {1\over 2 \sigma^2}) - 2\log\sigma \right.\right.\nonumber\\
&  &\left.\left. + {\rm F_1}\left[0,\frac{1}{2},-\frac{-\chi'^2}{2 \sigma ^2}\right]\right)
 + \sqrt{\frac{2}{\pi }} \chi'
 {\rm F_1}\left[1,\frac{3}{2},\frac{\chi'^2}{2 \sigma ^2}\right]\right\}
\end{eqnarray}
Here $\rm F_1$ is the hyper-geometric function of first kind. $\gamma_E$ is the Euler gamma which has a value of 0.577.

$K_2$ is defined as follows
\begin{eqnarray}
 K_2= \int_{-\infty}^{\infty}\left({\left(y-y_0\over\gamma\right)}^2 -1\right)W_y=
 \int_{0}^{\infty}\left({\left(\log\chi-y_0\over\gamma\right)}^2 -1\right){1\over \sqrt{2\pi}\sigma}e^{-{(\chi-\chi')^2\over2\sigma^2}}
\end{eqnarray}

For analytically performing this integration we expand $\log\chi$ in terms of $\chi\over 2\sigma$ as,
\begin{eqnarray}
 \log\chi=\log{\chi\over 2\sigma}+\log(2\sigma)=-{3\over2}+{\chi\over\sigma}-{\chi^2\over8 \sigma^2}+\log(2\sigma)\; .
\end{eqnarray}
Using this we get,
\begin{eqnarray}
 K_2 & = & {1\over 128 \sqrt{2 \pi} \gamma^2 \sigma^4 {\chi'}} e^{-{{\chi'}^2\over 2 \sigma^2}}
 \left[{\chi'} \left\{\left(-64 (7 + 4 y_0) +
 e^{{\chi'}^2\over 2 \sigma^2} \sqrt{2\pi}(235 + 208 y_0 + 64 y_0^2 - 64 \gamma^2)\right)\sigma^4 - \right.\right.\nonumber\\
& &  2\left(-93 - 16 y_0 + 8 e^{{\chi'}^2\over 2 \sigma^2} \sqrt{2 \pi} (15 + 8 y_0)\right) \sigma^3 {\chi'} +
    2 \left(-16 + e^{{\chi'}^2\over 2\sigma^2} \sqrt{2 \pi} (47 + 8 y_0)\right) \sigma^2 {\chi'}^2 - \nonumber \\
& &\left.\left.\left. 2 \left(-1 + 8 e^{{\chi'}^2\over 2 \sigma^2} \sqrt{2 \pi}\right) \sigma {\chi'}^3 +
    e^{{\chi'}^2\over 2 \sigma^2} \sqrt{2 \pi} {\chi'}^4\right\} + e^{{\chi'}^2\over 2 \sigma^2} \sqrt{2 \pi}
     {\chi'} \big((235 + 208 y_0 + 64 y_0^2 - 64 \gamma^2) \sigma^4 -\right.\right.\nonumber \\
& & \left.\left. 16 (15 + 8 y_0) \sigma^3 {\chi'} +
      2 (47 + 8 y_0) \sigma^2 {\chi'}^2 -
     16 \sigma {\chi'}^3 + {\chi'}^4\big) {\rm Erf}\left[{{\chi'}\over
     \sqrt{2} \sigma}\right] - \right.\right.\nonumber \\
& &   16 \sigma^2 \left\{{\chi'} \left((-16 +
            e^{{\chi'}^2\over 2\sigma^2} \sqrt{
             2 \pi} (13 + 8 y_0)) \sigma^2 -
         2 (-1 + 4 e^{{\chi'}^2\over 2\sigma^2} \sqrt{
             2 \pi}) \sigma {\chi'} +
         e^{{\chi'}^2\over 2\sigma^2} \sqrt{2 \pi} {\chi'}^2\right) +\right.\nonumber \\
&&\left.      e^{{\chi'}^2\over 2 \sigma^2} \sqrt{2 \pi}
        {\chi'} ((13 + 8 y_0) \sigma^2 -
         8 \sigma {\chi'} + {\chi'}^2) {\rm Erf}\left[{{\chi'}\over
        \sqrt{2} \sigma}\right]\right\} \log[2 \sigma] +\nonumber \\
&& \left.   64 e^{{\chi'}^2\over 2\sigma^2} \sqrt{
    2 \pi} \sigma^4 \left({\chi'} +
      {\chi'} {\rm Erf}\left[{{\chi'}\over \sqrt{2} \sigma}\right]\right) \log[
     2 \sigma]^2\right]
\end{eqnarray}

For $\sigma = 8.38\times10^{-8}$ the value of $K_1$ can be approximated as
$-0.458 + 4.15\times10^{12}{\chi'}^2$ and the value of $K_2$ can be approximated as 
$-0.461 + 1.6\times 10^{13}{\chi'}^2$. 
In the integration of $\delta N(\chi_i)$ (see \Eqn{deltaNR}), for ${g^2\over\lambda} = 2$ case, 
we have
\begin{eqnarray}
C_1 &=& -3.96\times10^{4} \left(1+ {K_1y_0\over\gamma} +
{K_2y_0^2\over 2\gamma^4} - {K_2\over 2\gamma^2}\right)\; , \nonumber \\
C_2 &=& -1.28\times10^{4}\left(1 + {K_1y_0\over\gamma} +
{K_2y_0^2\over 2\gamma^4} - {K_2\over 2\gamma^2}\right)
+ 3.96\times10^{4} \left({K_1\over\gamma} + {K_2y_0\over \gamma^4}\right)\; .
\end{eqnarray}

\vskip 1cm
\section*{Acknowledgement}
We would like to thank Gary Felder for some discussions regarding LATTICEEASY code. AM would like to acknowledge Palash B. Pal for fruitful
suggestions in deriving the effect of node shift on Lyapunov theorem. The authors are thankful to Koushik Dutta and Palash B. Pal for helping
in preparation of this manuscript. The authors acknowledge the Department of Atomic Energy (DAE, Govt. of India) for financial
assistance. 

\bibliographystyle{JHEP}
\bibliography{New-tex.bib}

\end{document}